\documentclass[10pt,twocolumn,letterpaper]{article}

\usepackage{cvpr}              % To produce the CAMERA-READY version

\usepackage{graphicx}
\usepackage{amsmath}
\usepackage{amssymb}
\usepackage{booktabs}
\usepackage{multirow}
\usepackage{array}
\usepackage[pagebackref,breaklinks,colorlinks]{hyperref}
\usepackage[dvipsnames]{xcolor}

\usepackage[capitalize]{cleveref}
\crefname{section}{Sec.}{Secs.}
\Crefname{section}{Section}{Sections}
\Crefname{table}{Table}{Tables}
\crefname{table}{Tab.}{Tabs.}

\begin{document}

%%%%%%%%% TITLE
\title{Classification of Brain Tumors using Hybrid Deep Learning Models}

\author{Neerav Nemchand Gala \\
Georgia Institute of Technology\\
{\tt\small ngala8@gatech.edu}
}
\maketitle

%%%%%%%%% ABSTRACT
\begin{abstract}
    The use of Convolutional Neural Networks (CNNs) has greatly improved the interpretation of medical images. However, conventional CNNs typically demand extensive computational resources and large training datasets. To address these limitations, this study applied transfer learning to achieve strong classification performance using fewer training samples. Specifically, the study compared EfficientNetV2 with its predecessor, EfficientNet, and with ResNet50 in classifying brain tumors into three types: glioma, meningioma, and pituitary tumors. Results showed that EfficientNetV2 delivered superior performance compared to the other models. However, this improvement came at the cost of increased training time, likely due to the model’s greater complexity.
\end{abstract}

%%%%%%%%% BODY TEXT
\section{Introduction}

According to the World Health Organization (WHO), in 2022, there were approximately 20 million new cancer cases and 9.7 million deaths \cite{WHO_2022_GlobalCancerBurden}. Brain tumors are a serious and prevalent health issue that can drastically impact life expectancy across all ages and genders. They result from the abnormal and uncontrolled growth of brain cells. An individual's prognosis depends on various factors, including the tumor's type, size, location, and the availability of effective treatment. Early detection and awareness of risk factors are crucial for successful management. Medical imaging, particularly Magnetic Resonance Imaging (MRI), plays a key role in the early detection and diagnosis of brain tumors. \cite{babu2023detection}.

Deep learning is a specialized branch of machine learning that gains its strength and flexibility by modeling the world as a layered hierarchy of concepts. In this structure, complex ideas are built upon simpler ones, with higher-level, more abstract representations derived from lower-level, less abstract features \cite{Goodfellow-et-al-2016}.

\begin{figure}[ht]
  \centering
  \includegraphics[width=0.45\textwidth]{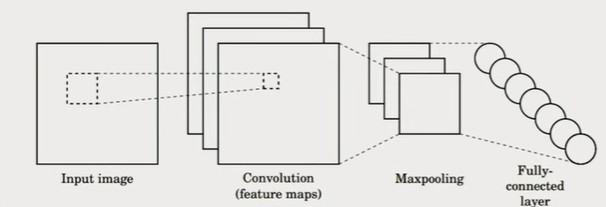}
  \caption{Simple CNN Structure\cite{amini2025introtodeeplearning}}
  \label{fig:basic_cnn}
\end{figure}

Deep learning has achieved significant success across a wide range of computer vision tasks, largely due to the effectiveness of convolutional neural networks (CNNs) that can preserve spatial information by extracting features from an image by applying convolution as shown in figure \ref{fig:basic_cnn}.

Convolutional Neural Networks (CNNs) have proven to be highly effective in medical image analysis, offering both precision and computational efficiency. A major benefit of CNNs is their ability to automatically extract meaningful features from images, which reduces the need for manual feature selection—a common requirement in traditional machine learning approaches. Despite their strengths, CNNs also present challenges. They generally depend on large, labeled datasets to achieve high accuracy, which can be difficult to obtain in medical contexts due to the time, cost, and expertise required for annotation. Furthermore, CNNs are computationally demanding, often requiring high-performance GPUs and substantial RAM for effective training and deployment.\cite{babu2023detection}.

To mitigate the limitations associated with traditional CNNs, this study employed transfer learning to improve performance using fewer training samples and to reduce training time. The research evaluates the effectiveness of the pretrained EfficientNetV2 model in classifying brain tumors into three categories—glioma, meningioma, and pituitary tumors. Its performance is compared against both its earlier version, EfficientNet, and another well-established pretrained CNN model, ResNet50.

\section{Related work}

\subsection{Deep Learning in Medical Imaging}

Recent advances in deep learning have significantly enhanced the automatic detection and classification of brain tumors from MRI scans. Traditional radiological diagnosis is time-intensive and requires substantial expertise, motivating the adoption of convolutional neural networks (CNNs) for automated interpretation. Studies over the past five years have established CNNs as the leading approach for this task, with research focusing on both accuracy and computational efficiency.

Siddique et al.~\cite{siddique2020robust} demonstrated the effectiveness of custom deep CNNs, achieving approximately 96\% accuracy on binary MRI tumor classification. Similarly, Khan et al.~\cite{khan2020brain} explored transfer learning with VGG-16, ResNet-50, and Inception-v3, reporting near-perfect accuracies, though on limited datasets. Moving beyond conventional architectures, Mzoughi et al.~\cite{mzoughi2020deep} leveraged a 3D CNN for glioma grading, emphasizing the value of volumetric data integration.

More recent studies have focused on advanced architectures. Nayak et al.~\cite{nayak2022efficientnet} employed EfficientNet to classify three types of brain tumors, achieving around 99\% accuracy, demonstrating its balance between depth and parameter efficiency. Raza et al.~\cite{raza2022deeptumornet} proposed DeepTumorNet, an enhanced GoogLeNet variant with additional layers, reaching 99.7\% accuracy. 

Vimala et al. \cite{babu2023detection} utilized transfer learning with five variations of pre-trained EfficientNets (EfficientNetB0 through EfficientNetB4) to perform multi-class classification of brain tumors using MR images and concluded that pre-trained EfficientNetB2 achieved impressive overall test accuracy of 99.06\%. Park and Kim~\cite{park2024explainable} incorporated EfficientNetV2L in their comparative study alongside ResNet and Vision Transformers, highlighting EfficientNetV2’s competitive performance and improved explainability using LIME and SHAP.

However, while EfficientNetV2 has shown promise, systematic evaluations againsts its predecessor on classifying brain MRI datasets remain limited. This gap underpins the motivation for the present project.

\subsection{Classification Models}

\subsubsection{ResNet50}
Since AlexNet's success in the 2012 ImageNet competition, deeper neural networks have been used to improve accuracy. However, as network depth increases, issues like vanishing or exploding gradients can arise, causing gradients to become too small or too large. This leads to higher training and test error rates, limiting the benefits of deeper architectures\cite{geeksforgeeks_resnet_2025}.

ResNet (Residual Neural Network) addresses the vanishing gradient problem by incorporating residual connections, allowing layers to learn modifications to their inputs rather than entirely new representations. This approach enables the successful training of very deep neural networks and has become a cornerstone in the field of image analysis, including medical imaging. By using identity-based skip connections, ResNet can bypass certain convolutional layers when minimal transformation is needed, improving gradient flow and reducing the risk of overfitting by avoiding unnecessary computations. \cite{resnet}. In particular, ResNet50 has 50 convolutional layers.

\subsubsection{EfficientNet}

EfficientNet is a new efficient network proposed by Google. Powered by an innovative compound scaling method that balances network depth, width, and image resolution for better accuracy at a fixed resource constraint. EfficientNet models can be scaled up very effectively, surpassing state-of-the-art accuracy with an order of magnitude fewer parameters and Floating-Point Operations Per Second (FLOPS) \cite{tan2020efficientnetrethinkingmodelscaling}. Due to these benefits, these models have become a common choice in the recent past for medical image analysis. 

\subsubsection{EfficientNetV2}

In 2021, EfficientNetsV2 models were released, that combine compound scaling method along with training-aware neural architecture search (NAS) to improve both
training speed and parameter efficiency. According to the creators of the model, EfficientNetV2 trains up to 11x faster while being up to 6.8x smaller than EfficientNet. \cite{tan2021efficientnetv2smallermodelsfaster}.

\section{Method}
\subsection{Data}
\begin{figure}[ht]
  \centering
  \includegraphics[width=0.5\textwidth]{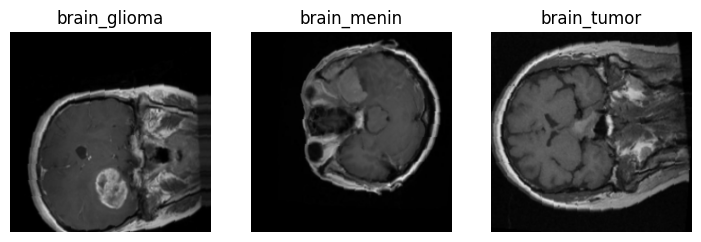}
  \captionsetup{justification=centering}
  \caption{Example of Dataset\\(Left: Glioma, Center: Meningioma, Right: Pituitary)}
  \label{fig:data}
\end{figure}

The dataset used for this project was the Bangladesh Brain Cancer MRI dataset \cite{rahman2024brain}. This open source dataset contains a total of 6056 images that are uniformly resized to 512x512 pixels. The images are systematically categorized into three distinct classes:
\begin{itemize}
    \item Brain Glioma: 2004 images
    \item Brain Menin: 2004 images
    \item Brain Tumor (pituitary\cite{kaggledatadiscussion}): 2048 images
\end{itemize} An example of a simple dataset is shown below (figure \ref{fig:data}).\\

The image preprocessing followed the requirements specified in the pretraining model, and the criteria for both models were the same: each image was resized to 224 x 224 pixels, Next, they were normalized with mean=[0.485, 0.456, 0.406], std=[0.229, 0.224, 0.225], which are the parameters of the pretrained ImageNet-1k dataset \cite{deng2009imagenet}.

The dataset was split into three parts: the training set consisted of 4245 images, the validation set of 607 images and the test set of 1214 images. While making the splits, the images were shuffled and stratified. 

\subsection{Data Augmentation Design}
The study followed the augmentation method of Gegenava \cite{geg_nikolas_92resnetv18_2025}, a related prior work. The details of the data augmentation are listed below:

\begin{itemize}
    \item Horizontally flip the given image randomly with a given probability $p=0.5$
    \item Vertically flip the given image randomly with a given probability $p=0.5$
    \item Randomly rotate a given image by a maximum angle of 15 degrees
    \item Cropping a random portion of image and resize it to 224 $\times$ 224 pixels
\end{itemize}

\subsection{Transfer Learning Setup}
For the study, the pretrained models listed in table \ref{tab:models} were considered:
\begin{table}[ht]
\centering
\caption{Pretrained Models}
\label{tab:models}
\begin{tabular}{p{3cm} p{4cm}}
\toprule
\multicolumn{1}{p{3cm}}{\textbf{Model Name}}       & \multicolumn{1}{p{4cm}}{\textbf{Description}} \\ 
\midrule
\multicolumn{1}{c}{tf\_efficientnetv2\_b0.in1k}   & A EfficientNet-v2 image classification model. Trained on ImageNet-1k in Tensorflow              \\ 
\multicolumn{1}{c}{tf\_efficientnet\_b0.in1k}       & A EfficientNet image classification model. Trained on ImageNet-1k in Tensorflow                          \\ 
\multicolumn{1}{c}{resnet50.a1\_in1k} & Trained on ImageNet-1k in timm               \\ 
\bottomrule
\end{tabular}
\end{table}
The pretrained models were taken from the timm module, which is deep-learning library created by Ross Wightman \cite{timm}. The models were retrained over 20 epochs, and the best model was determined based on the validation loss. All models were pre-trained on ImageNet, and after flattening the output of the model, the layers mentioned in table \ref{tab:network_layer} along with a softmax activation funciton were added to get the desired output.

\begin{table}[ht]
\centering
\caption{Modified Network Layers}
\label{tab:network_layer}
\begin{tabular}{>{\centering\arraybackslash}p{2cm} p{5.5cm}}
\toprule
\multicolumn{1}{c}{\textbf{Layer}} & \multicolumn{1}{c}{\textbf{Parameters}} \\ 
\midrule
\textbf{Linear} & in features=1280, out features=512, bias=True \\ 
\textbf{ReLU} & \\ 
\textbf{Dropout} & p=0.1, inplace=False \\ 
\textbf{Linear} & in features=512, out features=3, bias=True \\ 
\bottomrule
\end{tabular}
\end{table}

\subsection{Training Parameters}

\subsubsection{Cost Function}

The cost function used for this study was cross entropy loss and is given by:

\[
    J(\theta) = -\mathbb{E}_{\mathbf{x},y \sim \hat{p}_{\text{data}}} \log p_{\text{model}}(\mathbf{y} \mid \mathbf{x}).
\]

In other words, it is simply the negative log-likelihood, equivalently described as the cross-entropy ("distance") between the training data and the model distribution. 
The speciﬁc form of the cost function changes from model to model, depending on the speciﬁc form $\log p_{\text{model}}$ \cite{Goodfellow-et-al-2016}.

Cross entropy loss was one of the most common cost functions used to train neural networks as for classification.

\subsubsection{Optimizer}

The nonlinearity introduced by neural networks makes most interesting loss functions nonconvex. As a result, neural networks are typically trained using iterative, gradient-based optimization methods that aim to reduce the cost function to a very low value, rather than the exact solvers used for linear regression or the convex optimization techniques with global convergence guarantees employed for training logistic regression or SVMs \cite{Goodfellow-et-al-2016}.

For this study, the AdamW optimizer was used. It is an improved version of Adam that better handles weight decay (like shrinking weights to prevent overfitting)\cite{adamw}. It’s now the go-to optimizer for many deep learning tasks.

\subsubsection{Hyperparameters}

The hyperparameters were tuned using a subset of the data trained over 5 epochs. The hyperparameters in table \ref{tab:hyperparameters} were then implemented for the study:

\begin{table}[ht]
\centering
\caption{Training Parameters and Hyperparameters}
\label{tab:hyperparameters}
\begin{tabular}{>{\centering\arraybackslash}p{2cm} p{5.5cm}}
\toprule
\multicolumn{1}{c}{\textbf{Parameter}} & \multicolumn{1}{c}{\textbf{Hyperparameters}} \\ 
\midrule
Cross-Entropy Loss & label smoothing=0.1 \\ 
AdamW & initial learning rate=0.001 with 10 \% decrease every 3 epochs, weight decay=1e-4 \\ 
\bottomrule
\end{tabular}
\end{table}

\section{Experimental Setup}
\subsection{System Requirements}
The experiments were carried out on Google Colaboratory, an open-source notebook platform provided by Google. This platform offers access to both free and premium GPU and TPU resources, making it highly suitable for academic and research applications. The models were trained using NVIDIA T4 GPUs.

Python was used for programming and Pytorch for building and training the CNNs. The notebook used for this study can be found here \url{https://github.com/PyMyCode/BrainTumorClassification.git}

\subsection{Performance Metric Evaluation}
To evaluate and compare the performance of the CNN models, the following metrics were employed:

\begin{itemize}
    \item \textbf{Model Size}:\\
    The total number of trainable parameters in the network, which directly affects memory requirements and storage. It is typically reported as:
    \[
    \text{Model Size} = \sum_{l=1}^{L} \text{Parameters}_l
    \]
    where $L$ denotes the number of layers.

    \item \textbf{Training Time}:\\
    The total time required to train the model until convergence, measured in seconds, minutes, or hours.

    \item \textbf{Precision}:\\
    The proportion of correctly predicted positive observations to the total predicted positives:
    \[
    \text{Precision} = \frac{TP}{TP + FP}
    \]
    where $TP$ is the number of true positives and $FP$ is the number of false positives.

    \item \textbf{Recall (Sensitivity)}:\\
    The proportion of correctly predicted positive observations to all actual positives:
    \[
    \text{Recall} = \frac{TP}{TP + FN}
    \]
    where $FN$ is the number of false negatives.

    \item \textbf{F1-score}:\\
    The harmonic mean of precision and recall, providing a balance between the two:
    \[
    F1 = 2 \cdot \frac{\text{Precision} \times \text{Recall}}{\text{Precision} + \text{Recall}}
    \]

    \item \textbf{Accuracy}:\\
    The ratio of correctly predicted observations to the total observations:
    \[
    \text{Accuracy} = \frac{TP + TN}{TP + TN + FP + FN}
    \]
    where $TN$ denotes the number of true negatives.
\end{itemize}

\section{Experimental Results}
\subsection{ResNet50}

\begin{figure}[h]
  \centering
  \includegraphics[width=0.45\textwidth]{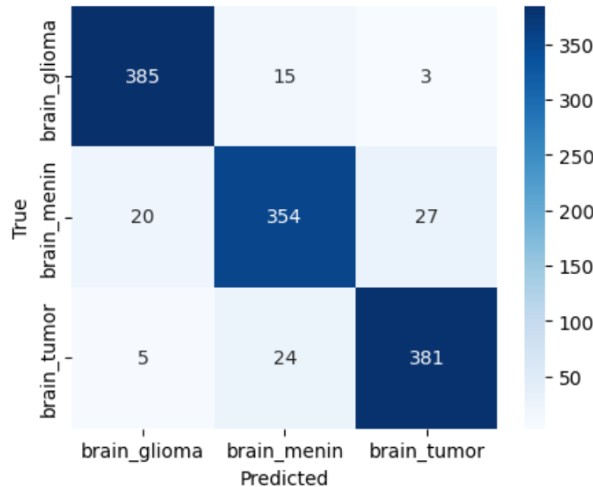}
  \captionsetup{justification=centering}
  \caption{ResNet50 Confusion Matrix}
  \label{fig:resnet50_confusionmatrix}
\end{figure}

This section displays the performance metrics for ResNet50 in table \ref{tab:resnet50_results} and figure \ref{fig:resnet50_confusionmatrix}.

\begin{table}[h]
    \centering
    \caption{ResNet50 Results}
    \begin{tabular}{lccc}
        \toprule
        & \textbf{Glioma} & \textbf{Meningioma} & \textbf{Tumor} \\
        \midrule
        \textbf{Precision} & 0.94 & 0.90 & 0.93 \\
        \textbf{Recall} & 0.96 & 0.88 & 0.93 \\
        \textbf{F1-score} & 0.95 & 0.89 & 0.93 \\
        \midrule
        \multicolumn{4}{l}{\textbf{Overall accuracy}: 0.92} \\
        \multicolumn{4}{l}{\textbf{Model size (in parameters)}: 24,558,659} \\
        \multicolumn{4}{l}{\textbf{Training time (in s)}: 1,074.56} \\
        \bottomrule
    \end{tabular}
    \label{tab:resnet50_results}
\end{table}

\subsection{EfficientNet}

\begin{figure}[h]
  \centering
  \includegraphics[width=0.45\textwidth]{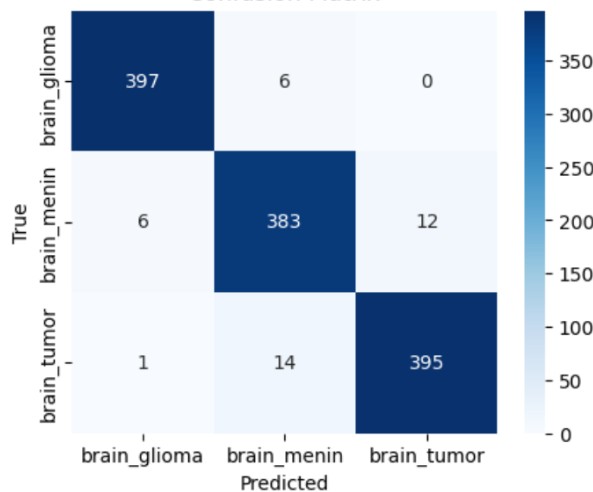}
  \captionsetup{justification=centering}
  \caption{EfficientNet Confusion Matrix}
  \label{fig:effnet_confusionmatrix}
\end{figure}

This section displays the performance metrics for ResNet50 in table \ref{tab:effnet_results} and figure \ref{fig:effnet_confusionmatrix}.

\begin{table}[h]
    \centering
    \caption{EfficientNet Results}
    \begin{tabular}{lccc}
        \toprule
        & \textbf{Glioma} & \textbf{Meningioma} & \textbf{Tumor} \\
        \midrule
        \textbf{Precision} & 0.98 & 0.95 & 0.97 \\
        \textbf{Recall} & 0.99 & 0.96 & 0.96 \\
        \textbf{F1-score} & 0.98 & 0.95 & 0.97 \\
        \midrule
        \multicolumn{4}{l}{\textbf{Overall accuracy}: 0.97} \\
        \multicolumn{4}{l}{\textbf{Model size (in parameters)}: 4,664,959} \\
        \multicolumn{4}{l}{\textbf{Training time (in s)}: 917.62} \\
        \bottomrule
    \end{tabular}
    \label{tab:effnet_results}
\end{table}

\subsection{EfficientNetV2}

\begin{figure}[h]
  \centering
  \includegraphics[width=0.45\textwidth]{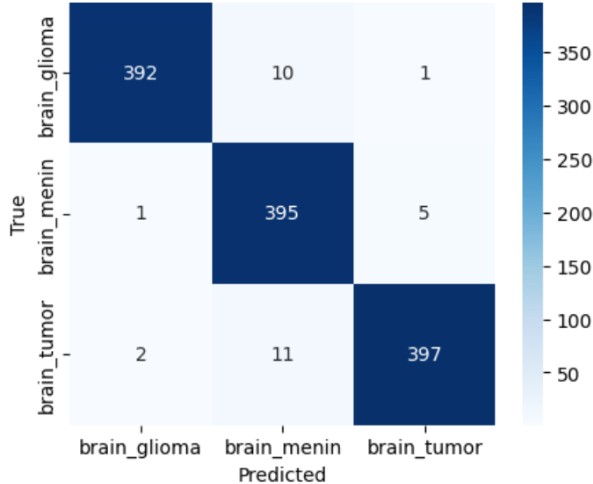}
  \captionsetup{justification=centering}
  \caption{EfficientNetV2 Confusion Matrix}
  \label{fig:effnetv2_confusionmatrix}
\end{figure}

This section displays the performance metrics for ResNet50 in table \ref{tab:effnetv2_results} and figure \ref{fig:effnetv2_confusionmatrix}.

\begin{table}[h]
    \centering
    \caption{EfficientNetV2 Results}
    \begin{tabular}{lccc}
        \toprule
        & \textbf{Glioma} & \textbf{Meningioma} & \textbf{Tumor} \\
        \midrule
        \textbf{Precision} & 0.99 & 0.95 & 0.99 \\
        \textbf{Recall} & 0.97 & 0.99 & 0.97 \\
        \textbf{F1-score} & 0.98 & 0.97 & 0.98 \\
        \midrule
        \multicolumn{4}{l}{\textbf{Overall accuracy}: 0.98} \\
        \multicolumn{4}{l}{\textbf{Model size (in parameters)}: 6,516,115} \\
        \multicolumn{4}{l}{\textbf{Training time (in s)}: 3,428.94} \\
        \bottomrule
    \end{tabular}
    \label{tab:effnetv2_results}
\end{table}

\section{Discussion}

Tables \ref{tab:resnet50_results}, \ref{tab:effnet_results} and \ref{tab:effnetv2_results} show that even though ResNet50 has more parameters ($\approx24.6$ million), its overall accuracy (0.92) on the test dataset is lesser than both EfficientNet (overall accuracy of 0.97 with $\approx 4.7$ million parameters) and EfficientNetV2 (overall accuracy of 0.98 with $\approx 6.5$ million parameters). In particular, ResNet50 was relatively poor in distinguishing meningioma from pituitary tumors (figure \ref{fig:resnet50_confusionmatrix}). 

In the EfficientNet family, EfficientNetV2 had a slightly better overall accuracy (0.98) in comparison to EfficientNet (0.97). In particular, EfficientNetV2 performed better at classifying meningioma and pituitary as compared to EfficientNet (refer to figure \ref{fig:effnet_confusionmatrix} and figure \ref{fig:effnetv2_confusionmatrix}).  

The training time of EfficientNetV2 (3,428.94 s) is longer than that of EfficientNet (917.62 s) and ResNet50 (1,074.56 s). This could be due to the higher complexity of the EfficientNetV2 model. 

\section{Conclusion}

By combine compound scaling method along with training aware neural architecture search (NAS), EfficientNetV2 performs significantly better than the benchmark ResNet50 and slightly better than its predecessor, EfficientNet. However, this higher performance comes with longer training time, which could be due to it's complexity.

%%%%%%%%% REFERENCES
{\small
\bibliographystyle{ieee_fullname}
\bibliography{ref}
}

\end{document}